# Honeycomb monolithic design to enhance the performance of Ni-based catalysts for dry reforming of methane

Fazia Agueniou, Hilario Vidal *, M. Pilar Yeste, Juan C. Hernández-Garrido, Miguel A. Cauqui, José M. Rodríguez-Izquierdo, José J. Calvino, José M. Gatica

Departamento de Ciencia de Materiales e Ingeniería Metalúrgica y Química Inorgánica, e IMEYMAT, Instituto Universitario de Investigación en Microscopía Electrónica y Materiales, Universidad de Cádiz, 11510 Puerto Real, Spain.

*Corresponding author

Tel. +34-956-012744

E-mail address: hilario.vidal@uca.es (Hilario Vidal)

# ABSTRACT

Supported Ni catalysts (4.5 wt%) using a Ce-Zr oxide (18/82 molar ratio and a ceria-rich surface) depicting advanced redox properties, were deposited by washcoating over cordierite honeycombs (230 and 400 cpsi). FIB-STEM unveiled nanostructure details otherwise undistinguishable by conventional techniques. The catalytic performance was evaluated in the dry reforming of methane at 700-900 ºC, using a $CH_4:CO_2$ 1:1 feedstock, and exploring high Weight Hourly Space Velocity (115-346 L $g^{-1}$ $h^{-1}$). The structured catalysts exhibited better performance than the corresponding powders, reaching values close to thermodynamic limits for both reactants conversion and $H_2/CO$ ratio, from 750 ºC, and no deactivation was observed in prolonged experiments (24-48 hours). This was related to both the high catalyst efficiency after being deposited with low loading on the cordierite and the intrinsic advantages of the monolithic reactor, like preventing from the kinetic control that operates in powdered samples under high WHSV or limiting the deactivation.

# 1. Introduction

Dry Reforming of Methane (DRM) is most likely one of the reactions that has attracted more attention in the last decade in the field of heterogeneous catalysis. This is due to the growing interest in syngas production processes [1], and the simultaneous abatement of two significant greenhouse gases such as $CO_2$ and $CH_4$ [2].

Although noble metals have been also employed [3], undoubtedly nickel-based catalysts have become the best option for this reaction because of their lower cost while keeping a high performance [4,5]. Nevertheless, it is well-known the tendency of nickel to suffer deactivation due to carbonaceous deposits originated in the reaction onto the metal surface. Among others, the appropriate selection of the support has been one of the most promising strategies to increase the stability of these catalysts [6]. Thus, for example, nickel has been supported over Ce-Zr and Ce-Pr mixed oxides, whose improved redox properties allow the oxidation of the carbon accumulated onto the metal [7,8]. A recent study explored the use as DRM catalysts of nickel supported onto Ce-Zr oxides previously activated to enhance their redox properties by using a thermo-chemical protocol consisting on a severe reduction at 950 ºC followed by a mild oxidation at 500 ºC [9]. Some of these formulations showed high activity and stability against deactivation caused by carbon formation even under very severe operating conditions.

However, in most published works dealing with DRM, catalysts are used in the form of powder, granules or pellets [10-14], as well illustrated even in the most recent reviews [5]. In contrast, the use in this process of structured catalysts is, so far, very scarcely reported; the only few works available having used either cordierite [15-18], silicon carbide [18] or stainless steel [19] honeycombs and, more rarely, membrane reactors [20]. This is doubly surprising if the well-known advantages of such design are considered [21], and also considering the precedents in the use of honeycomb monoliths and foams in hydrocarbons reforming [22] and $CO_2$ methanation [23,24]. Moreover, in the case of DRM, the conventional operation in diffusion mode would be favoured when the catalyst consists in a thin coating of the ceramic walls. In addition, the open structure of the honeycomb could minimize the overpressure or even blocking phenomena produced by massive deposition of carbonaceous species, which eventually occur when using powdered-packed reactors.

It is also important to consider that in the above cited works in which honeycomb monoliths were employed for the DRM reaction, the studied structured catalyst was obtained by wet impregnation of the monolith with precursor salts [15,18] or sol-gel methods [18] and not by washcoating, which precludes any comparison between the catalyst as a powder and after deposition on the monolith. The same approach was followed in [16] where, additionally, a noble metal (Ru, 0.5% wt.) was used as promoter of nickel. Regarding the study performed by Kohn et al. [17], the authors used a commercial BASF catalyst (Rh/γ-$Al_2O_3$) as active phase loaded onto the monolith, but did not provide any detail about its preparation. Finally, in [19] Fukuhara et al. did not either consider the powder catalyst as a reference because the active phase, Ni doped with Pd and Sn, used to coat a stainless-steel honeycomb substrate was obtained by an electroless plating method.

Considering the background briefly revised above, the first purpose of this work is to give a step forward by depositing a nickel-based advanced formulation onto cordierite honeycombs using low specific loadings in order to maximize the advantages of the structured design [15], testing them in the DRM reaction and making a comparison with the same catalyst in the form of powder. Moreover, the monolithic design has allowed us exploring experimental conditions, in terms of weight hourly space velocities (WHSV), which have been rarely considered in previous works [5,25]. To rationalize the results, an in-depth structural study by both macroscopic and atomic scale techniques was carried out. Concerning the latter, a combination of Focused Ion Beam (FIB) electron microscopy sample preparations with further analysis by Scanning Transmission Electron Microscopy (STEM) has been used, previously proved as a quite powerful tool for the characterization at the finest, subnanometric, and atomic, scales of coated honeycomb monoliths and to understand their macroscopic performance [26,27]. Moreover, such studies revealed that using this experimental approach it is possible to detect microstructural features of the final catalytic device out of reach for other conventional characterization techniques. Therefore, a second goal of this work is applying this approach to investigate the prepared Ni-based honeycomb-type catalysts.

The work here presented deepens our preliminary communication [28] by widening the set of samples investigated and applying the needed physico-chemical techniques for a better understanding of the relationships between catalysts properties and performance. Now we have included new reference samples to pay special attention to the role played by the alumina employed for the structured catalysts preparation and the effect of the honeycomb substrate cell density. Regarding the techniques, we have incorporated results obtained by means of X-ray fluorescence, $N_2$ physisorption, X-Ray Diffraction, $H_2$ and $O_2$ volumetric adsorption, TPR, and Temperature-Programmed Reaction experiments followed by mass spectrometry. In this way, we can provide information at compositional, textural, structural and chemical level, which is necessary to explain the origin of the outstanding behaviour reported in our previous study [28].

## 2. Experimental

*2.1. Powdered catalyst*

The ceria-zirconia support was prepared from a commercial nano-sized zirconia from Tecnan-Nanomat S.L (78 m$^2$ g$^{-1}$ and 10-15 nm of BET and average particle size, respectively) which was impregnated to incipient wetness with a Ce(NO$_3$)$_3$·6H$_2$O (99.99%, Sigma Aldrich) aqueous solution (0.88 M) to obtain an oxide having a 20%Ce-80%Zr nominal molar composition. After drying in air (110 ºC, 12 h) and grinding (75 μm), the sample was calcined (500 ºC, 1 h). The resulting oxide was further activated for its redox properties enhancement following a thermo-chemical aging protocol known as SRMO (Severe Reduction-Mild Oxidation) proposed by our lab [29]. Briefly, it consists in a reduction treatment under H$_2$(5%)/Ar (950 ºC, 2 h), followed by He flushing (950 ºC, 1 h) and oxidation with pulses of O$_2$(5%)/He at room temperature, and finally heating under the oxidising mixture up to 500 ºC (1 h). In the following, the SRMO treated Ceria-Zirconia oxide will be referred as CZ.

The supported nickel catalyst, named NiCZ, was prepared by incipient wetness impregnation with a Ni(NO$_3$)$_2$·6H$_2$O (99.99%, Sigma Aldrich) aqueous solution (1.01 M) aiming to reach a nominal 5 wt% metal loading. The catalytic precursor was dried (110 ºC, 12 h), calcined (400 ºC, 1 h) and finally grinded (75 μm).

*2.2. Honeycomb monolithic catalysts*

Cordierite blocks (Corning) with a cell density of 36 cells cm$^{-2}$ (230 cpsi) or 62 cells cm$^{-2}$ (400 cpsi), and wall thickness of 0.18 mm (7 mil) or 0.17 mm (6.5 mil), respectively, were cut to obtain cylindrical pieces having 13 mm of diameter, 47 mm of length and 2 g (230 cpsi samples) or 2.5 g (400 cpsi samples) of approximate weight. Monolithic catalysts were prepared by washcoating from a slurry (stabilized at pH 4.0 using acetic acid) containing the NiCZ catalyst or the CZ support, following the methodology well described by Prof. Montes laboratory [30]. The granulometry ($d_{90}$ = 0.80 µm) and the Z potential (36 mV at pH = 4, with the isoelectric point at pH = 6.8) data for the NiCZ sample are proper for the washcoating method as revised for example in [30]. Therefore, we prepared slurries containing the NiCZ or the CZ sample (19.1 wt%), polyvinyl alcohol (1.7 wt%), Nyacol AL20 colloidal alumina (4.2 wt%) and water. As far as our objective was to reach low catalyst loadings, we decided to prepare slurries with low viscosity (5.1 mPa s). Cordierite pieces were immersed (3 cm min$^{-1}$) in this slurry, kept fully immersed for 90 s, the first 15 s under ultra-sonication. They were further pulled out at the same rate, and the excess of slurry was removed by air flowing. The pieces were then dried at 120 ºC for 30 min and submitted to one or two new coating/drying cycles until reaching the desired final specific loading, around 0.4 mg cm$^{-2}$ corresponding to a washcoat loading about 25 mg per g$^{-1}$ of support. Finally, all the monoliths were calcined (5 ºC min$^{-1}$) at 450 ºC (1 h).

The reached active phase loading was estimated from the weight gain after calcination, and its adherence (expressed in percentage) from the weight loss after immersion in petroleum ether under ultrasounds (30 min). The prepared monolithic catalysts were named as H230 or H400 (according to the cell density of the cordierite) followed by CZ or NiCZ (to denote the nature of the coating). The residual slurries, containing either CZ or NiCZ, were dried and submitted to calcination at 450 ºC (1 h) to obtain the CZ-S and NiCZ-S reference powdered samples, respectively.

*2.3. Catalysts characterization*

The prepared catalysts were characterized by ICP-AES using a Thermo Elemental Iris Intrepid equipment. Complementary compositional analyses were performed through X-ray micro-fluorescence (XRF) in a Bruker S4 Pioneer spectrophotometer. Granulometric distribution of the powdered materials dispersed in distilled water was obtained using a Malvern Mastersizer 2000 laser diffraction instrument while the Z potential measurements were carried out in a DLS Malvern Zetasizer Nano-ZS device. The viscosity of the slurries was measured by means of a concentric cylinder viscometer from Brookfield model DV-II+. Textural characterization (BET specific surface area and pore volume measurements) was carried out over pre-evacuated at 200 ºC (2 h) samples in a Quantachrome Autosorb IQ equipment. X-ray diffraction (XRD) diagrams were recorded for powdered samples employing a Bruker diffractometer and Cu Kα radiation. XRD data were processed using the Powder Cell 2.4 software in order to estimate phases percentages and average crystallite sizes. The catalysts were also studied by scanning electron microscopy (SEM), obtaining both images and Energy-dispersive X-ray (EDX) compositional analysis spectra in a FEG Nova NanoSEM 450 microscope operating at 30 kV. In addition, both $H_2$ and $O_2$ chemisorption volumetric isotherms were obtained at 50 ºC for samples which had been previously reduced with $H_2$(5%)/Ar at 600 ºC (2 h) and evacuated at 600 ºC (1 h). For these experiments, the adsorbed volumes were determined by extrapolation of the linear part of the adsorption isotherms (in the 100-300 Torr range) to zero pressure, and a chemisorption stoichiometry ratio of 1:1 was assumed for both molecule probes, H:Ni and O:Ni. Temperature-Programmed Reduction (TPR) diagrams and reaction profiles at programmed temperature were recorded in an Autochem II 2920 equipped with thermal conductivity detector (TCD) and a Pfeiffer QSM-200 mass spectrometer, respectively.

Sample preparation for STEM observation was carried out in a Zeiss Auriga FIB-SEM system operating at 30 keV (ion beam) and equipped with an Omniprobe micromanipulator, which allows an in-situ lift-out of the electron-transparent lamellas. Previous to the FIB preparation, the sample was covered by sputtering (ex situ) with a thin layer (<10 nm) of Au to avoid as much as possible charge effects during the preparation process. A protective layer (in this case, platinum layer: 20 × 2 × 2 μm) was placed in situ via ion induced deposition on the specific region of interest in order to avoid severe damage on the surface. STEM studies, both in High Angle-Annular Dark-Field (HAADF) and analytical EDX modes were performed in an aberration-corrected FEI Titan[3] Themis 60-300 microscope operating at 200 kV. A condenser aperture of 50 μm and a 91 mm camera length was used, obtaining an electron probe with a convergence angle of 20 mrad. In order to get a high signal-to-noise ratio, a beam current of 0.2 nA was used. The EDX hypermaps were recorded using a Super-X EDX system, which gathers the signals from 4 window-less EDX detectors surrounding the TEM sample and provides collection of the signals from a solid angle close to 1 srad.

*2.4.  Catalytic tests*

Catalytic performance in the DRM reaction was evaluated for both powdered and honeycomb monolithic samples. These tests were run in quartz reactors, at atmospheric pressure and using a 1:1 mixture of pure $CH_4$ and $CO_2$ as feedstock. For powders, 26 mg of sample were diluted in 52 mg of SiC, adjusting the total flow to 50 mL min$^{-1}$. In the case of monoliths, we employed 21 mm long pieces, containing 13 mg of catalyst, and a total flow of 25, 50 or 75 mL min$^{-1}$ depending on the WHSV selected. Small pieces of quartz at the inlet were included to guarantee a turbulent flow. The investigated range of WHSV, expressed as the ratio between reactants flow and sample amount, was 115-346 L g$^{-1}$ h$^{-1}$, while Time on Stream (TOS) in these experiments varied from 24 to 48 h. In all cases the catalysts were subjected to a reduction pre-treatment with 60 mL min$^{-1}$ of $H_2$(5%)/Ar at 600 ºC (2 h). The reaction temperature was measured by means of a thermocouple located in contact with the quartz reactor at the position of the catalytic bed. The gas analysis at the inlet and outlet of the reactor was performed by gas chromatography (Bruker 450-GC), using helium (25 mL min$^{-1}$) as carrier inert gas.

Reactants ($CH_4$ and $CO_2$) conversion values were calculated from the inlet and outlet molar fractions of the individual gases as follows:

$$CH_4 \text{ Conversion (\%)} = 100 \times \frac{[CH_4]_{inlet} - [CH_4]_{outlet}}{[CH_4]_{inlet}}$$

$$CO_2 \text{ Conversion (\%)} = 100 \times \frac{[CO_2]_{inlet} - [CO_2]_{outlet}}{[CO_2]_{inlet}}$$

The estimate of thermodynamic conversion limit values was performed using the DETCHEM software [31].

A complementary study aimed to establishing the temperature range of the above commented catalytic tests were done by means of Temperature-Programmed Reaction experiments, heating (5 °C min$^{-1}$) the samples under a flow of $CH_4$(20%)/$CO_2$(20%)/He and using mass spectrometry as analytical tool.

## 3. Results and discussion

*3.1. Characterization and properties of the powdered samples*

The Ce/Zr molar ratio in the CZ sample and the nickel content in the NiCZ powdered catalyst (Table 1) were close to the nominal values. The nickel content of the powdered catalyst proportionally decreased in the NiCZ-S sample due to mixing with the alumina used as additive to prepare stable slurries of the active phases prior to the cordierite coating. This alumina is also responsible of the increase of BET specific surface area and pore volume observed for both the support and the catalyst (illustrated in Table 1 as well).

**Table 1.** Physico-chemical properties of the powdered samples as measured by means of ICP-AES, N$_2$ physisorption and volumetric chemisorption**.**

|  | CZ | CZ-S | NiCZ | NiCZ-S |
|---|---|---|---|---|
| Ce/Zr (molar ratio) | 0.18 | 0.18 | 0.18 | 0.18 |
| Ni (wt%) | n.a. | n.a. | 4.5 | 3.9 |
| $S_{BET}$ (m$^2$ g$^{-1}$) | 13.3 | 23.0 | 15.6 | 20.7 |
| $V_p$ (cm$^3$ g$^{-1}$) | 0.051 | 0.074 | 0.048 | 0.064 |
| H$_2$ adsorption (µmol g$^{-1}$) [a] | - | - | 21.9 | 7.0 |
| O$_2$ adsorption (mmol g$^{-1}$) [a] | 0.216 | 0.221 | 0.227 | 0.238 |

[a] Data obtained for samples activated with H$_2$(5%)/Ar at 600 ºC.

It is worth mentioning that all studied samples were mesoporous, as denoted by both the N$_2$ adsorption-desorption isotherms and the derived pore size distribution curves (Fig. 1). In all cases the isotherms were of type IV with a H1 hysteresis loop, usually associated with porous materials that consist of agglomerates or compacts of approximately uniform spheres in fairly regular array, and hence having relatively narrow pore size distributions [32]. On the other hand, the pore size distribution curves show that, as expected, incorporating nickel to the CZ support induces the disappearance of some pores, especially those above 10 nm in diameter, most likely due to partial blocking by the nanostructures comprising this metal component. A similar effect is observed when comparing the curves of CZ-S and NiCZ-S, as it was also previously reported for other supported metal catalysts [30,33]. In contrast, the increase of porosity due to the addition of alumina, as commented above, becomes evident when comparing the curves of the CZ - CZ-S or NiCZ - NiCZ-S counterparts.

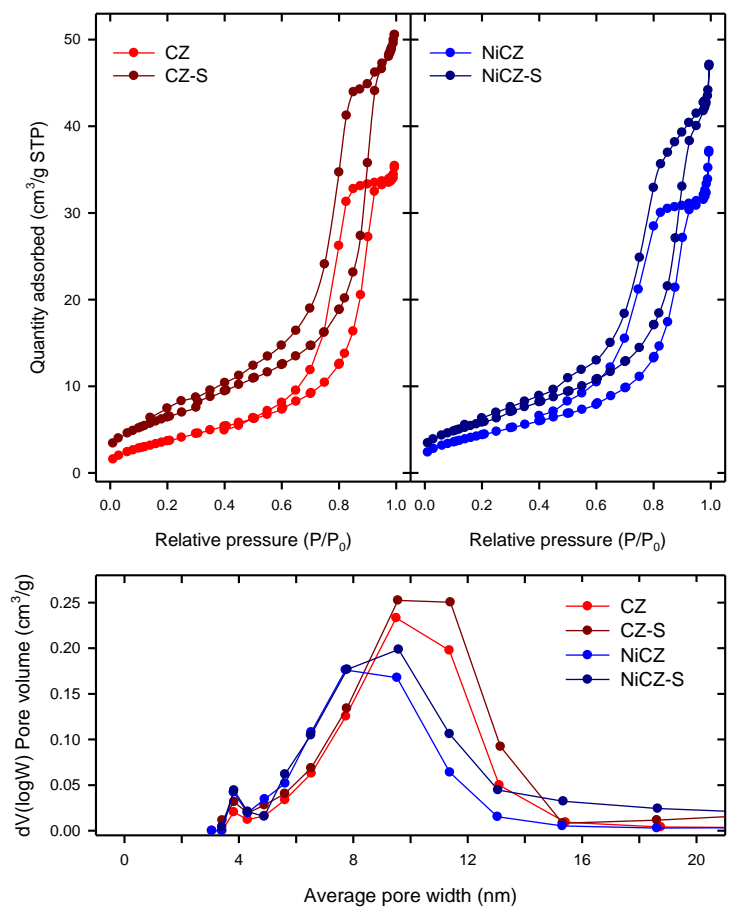

**Fig. 1.** N$_2$ physisorption isotherms (upper part) obtained for the powdered supports and catalysts samples of this study. The full and empty symbols correspond to adsorption and desorption branches respectively. Pore size distribution curves obtained from the N$_2$ physisorption experiments are included below.

Fig. 2 illustrates the effect on the structure of incorporating cerium onto the commercial zirconium oxide and applying the SRMO treatment proposed in [29] to improve the redox performance of the catalytic support, according to XRD. For comparison, the diffractograms of a pure ZrO$_2$ sample both fresh and after the same SRMO treatment are also included. The starting zirconia (36% monoclinic and 64% tetragonal) suffers the martensitic tetragonal-to-monoclinic phase transformation [34] when the SRMO is applied, rendering a 91% monoclinic and 9% tetragonal material. The presence of cerium favors the cubic/tetragonal phases [35] and indeed our CZ sample contains only 10% monoclinic (JCPDS 37-1484), 5% cubic (JCPDS 27-0997) and 85% tetragonal (JCPDS 80-0965) zirconia phases. The absence of peaks related to pure ceria and the presence of shoulders located around 29.5º, 49.4º and 58.3º besides the main diffraction peaks suggests the formation of a ceria-zirconia solid solution that can be related with a tetragonal phase referred to as $Ce_{0.18}Zr_{0.82}O_2$ (JCPDS 80-0785).

In the case of the nickel catalyst, relatively low intense peaks, which can be attributed to the (200) and (111) planes of monoclinic nickel oxide (JCPDS 65-6920), were additionally detected. The NiO crystal size estimated by the Debye-Scherrer formula was 17 nm and 15 nm for the NiCZ and NiCZ-S samples, respectively. Regarding the samples containing alumina, practically overlapping XRD diagrams were recorded when comparing CZ with CZ-S, and NiCZ with NiCZ-S, respectively. These results showed that the massive structural nature of the materials is not modified in the suspension used for the preparation of the monolithic catalysts.

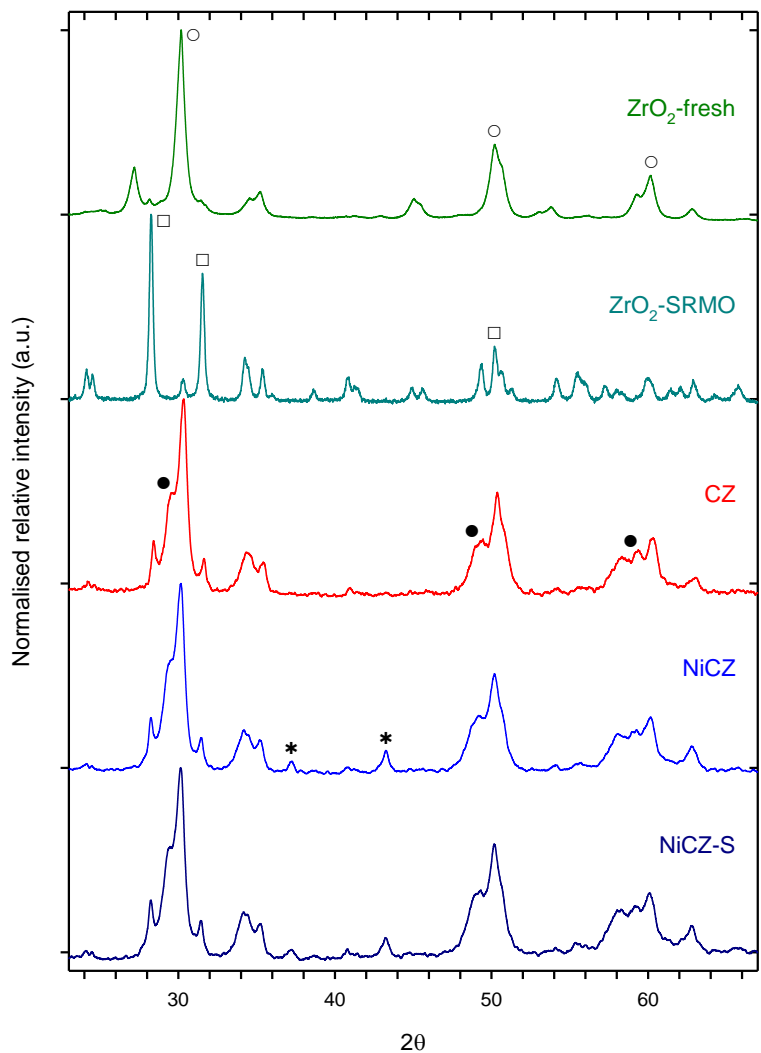

**Fig. 2.** XRD diagrams obtained from the powdered samples including the zirconia used in this study as received and submitted to the ageing treatment employed in the case of ceria-containing samples. Main diffraction peaks corresponding to monoclinic JCPDS 37-1484 (□) and tetragonal JCPDS 80-0965 (○) zirconia; tetragonal $Ce_{0.18}Zr_{0.82}O_2$ JCPDS 80-0785 (●); and monoclinic NiO JCPDS 65-6920 (✱), are indicated.

The TPR results (Fig. 3) indicated that cerium reduction in CZ sample starts at 250 ºC, reaching maxima at 480 and 520 ºC resulting from a non-symmetric and broad profile. Considering that temperature maxima around 600 ºC are reported for the main reduction peaks in similar Ce-poor/Zr-rich mixed oxides [36], these results probe the effectiveness of the SRMO pretreatment aging protocol [29] for the oxide support that we have employed. The NiCZ profile (Fig. 3) is quite complex as shows an intense maximum at 325 ºC and shoulders located around 175, 260, 290 and 360 ºC, with a very low but continuous hydrogen consumption up to 750 ºC. The quantitative analysis from the signals (also included in Fig. 3) is consistent with the reduction towards Ce(III) and Ni(0) of the cerium and nickel total amounts for all the samples, considering their initial state in the form of Ce(IV) and Ni(II).

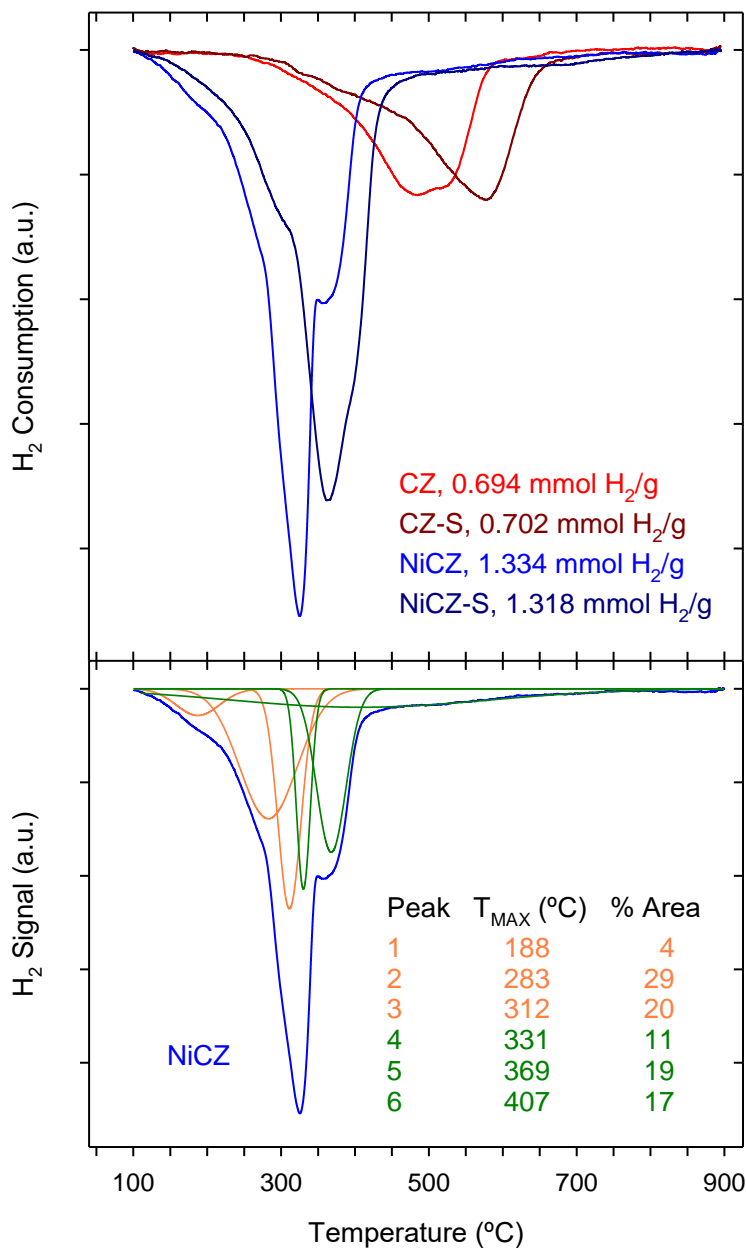

**Fig. 3.** TPR profiles of the powdered samples under $H_2(5\%)Ar$ and a heating rate of 10 ºC min$^{-1}$, including quantitative data (above) and deconvoluted patterns with the position of maxima and area percentages (below) for the NiCZ sample.

The detailed interpretation of TPR profiles of nickel supported on ceria-zirconia binary oxides is difficult [36] as far as there is no agreement in literature regarding the ascription of their features. TPR peaks occurring from 150 ºC and with maxima in the 250-400 ºC temperature range have been reported by using deconvolution procedures [9,36], and the differences in temperature for the reduction of supported NiO have been explained in terms of different distribution of particle sizes and/or degrees of interaction with the support [37,38]. On the other hand, reduction of the support and the nickel species can occur simultaneously, promoted by spillover of hydrogen species [9,39], thus significantly lowering the temperature at which the ceria-zirconia solid solution is reduced. Taking into account the quantitative data, and applying the data treatment previously reported [9,36], we can tentatively attribute the first part of the TPR patterns in the NiCZ sample (namely, that fitted by the first three peaks) to the reduction of nickel species, and the remaining high temperature part to the reduction of the support (Fig. 3). Indeed, the hydrogen consumption corresponding to the first three peaks in the deconvoluted patterns for the NiCZ sample (0.707 mmol $H_2$/g) is close to that theoretically required for complete reduction of nickel from Ni(II) to Ni(0) (0.767 mmol $H_2$/g). On the other hand, the hydrogen consumption corresponding to the last three peaks (0.627 mmol $H_2$/g) fits well to that theoretically required for the reduction of cerium from Ce(IV) to Ce(III) (0.644 mmol $H_2$/g). In the case of the CZ-S and NiCZ-S alumina-containing samples, which are the best references of the phases deposited on the monolithic catalysts, the recorded TPR curves suggest some chemical interaction between the nickel and the support oxide with the alumina; this can be inferred from the shift of the maxima of the reduction peaks towards higher temperatures with respect to the CZ and NiCZ samples (580 ºC and 365 ºC for CZ-S and NiCZ-S, respectively). This is in good agreement with previous reports evidencing that the presence of alumina in nickel supported onto ceria-zirconia mixed oxides shifts to higher temperatures the reduction of both NiO [40,41] and ceria-containing phases [42,43].

The results commented above allowed selecting 600 ºC as the catalysts activation temperature in this study, as far as this temperature guaranties that the entire nickel and most of the cerium contained in the samples should be in a reduced form. Previous studies also indicated the need of reducing supported nickel catalysts at relatively high temperatures (> 550 ºC) to guarantee total reduction of this metal [44], a prerequisite well-stablished for obtaining the better catalytic results in DRM [45,46].

The nickel catalysts, activated at 600 ºC, chemisorbed a very low amount of hydrogen (Table 1), rendering H/Ni ratios of 0.06 and 0.02 for the NiCZ and NiCZ-S samples, respectively. Similar results, that could be interpreted as evidences of very low nickel dispersion, have been related instead to the existence of strong metal-support interaction (SMSI) that hinders the hydrogen adsorption in ceria-zirconia supported Ni catalysts when submitted to reduction treatments at 600 ºC [7]. In fact, this effect was early reported for nickel catalysts supported onto alumina, silica and titania [47]. The use of $O_2$ as molecule probe, as proposed in [48], can provide adequate nickel dispersion values when the oxygen consumption from the support is taken as reference (Table 1). The resulting Ni dispersion percentages were 6 (O/Ni = 0.06) and 10 (O/Ni = 0.10) for NiCZ and NiCZ-S catalysts, respectively. The use of the relationship that correlates the average nickel particle size and the metal dispersion in nickel supported catalyst [47] renders 17 and 10 nm for NiCZ and NiCZ-S samples, respectively, in reasonable agreement with the results estimated from XRD line broadening once the structural correction from the reduction of NiO to Ni phases is applied (14 and 13 nm for NiCZ and NiCZ-S, respectively).

*3.2. Honeycomb catalysts characterization*

As indicated in the experimental section, the NiCZ was deposited over cordierite honeycombs by washcoating, obtaining relatively low catalysts loadings: 0.33, 0.36, 0.41 and 0.38 mg cm$^{-2}$ for H230CZ, H230NiCZ, H400CZ and H400NiCZ samples, respectively. It should be highlighted that these values are at least one order of magnitude higher in conventional monoliths [49]. The adherence of the coatings was evaluated from the washcoating material weight that remained attached to the cordierite after the ultrasound treatment test, resulting 91, 92, 97 and 95% for H230CZ, H230NiCZ, H400CZ and H400NiCZ samples, respectively. These values are acceptable considering that a maximum weight loss limit of 10% has been obtained for oxide phases-washcoated cordierite honeycomb monoliths under optimum conditions [50]. It is also remarkable that these parameters were not reported in the few previous works dealing with honeycomb-type catalysts for DRM [15,16,18,19], except one in which a cordierite honeycomb was washcoated with a Rh/γ-Al$_2$O$_3$ [17], but the authors did not provide the adherence of the coating and its loading was much higher (43.8 mg cm$^{-2}$).

The use of SEM with EDX compositional analysis at the micron scale allowed detecting the phase deposited over the cordierite walls. For illustrative purposes, Fig. 4A shows a wall region of the H230NiCZ monolith, with an area of 150 x 125 μm$^2$, in which the presence of Ni, Ce and Zr is clearly observed. Note how the particles of the catalyst are scattered over the surface of the monolith walls, in a situation far from that corresponding to a continuous film. Likewise, X-ray micro-fluorescence data (Fig. 4B) evidenced a partial coating of the cordierite walls by the active phase in massive state, in good agreement with both the low load of the prepared honeycomb monolithic catalysts and the EDX maps just commented. Moreover, both techniques suggest an appropriate Ni-Ce-Zr interaction in the zones where the deposits of catalyst are found, since the signals of the three elements are present at the same or nearby locations.

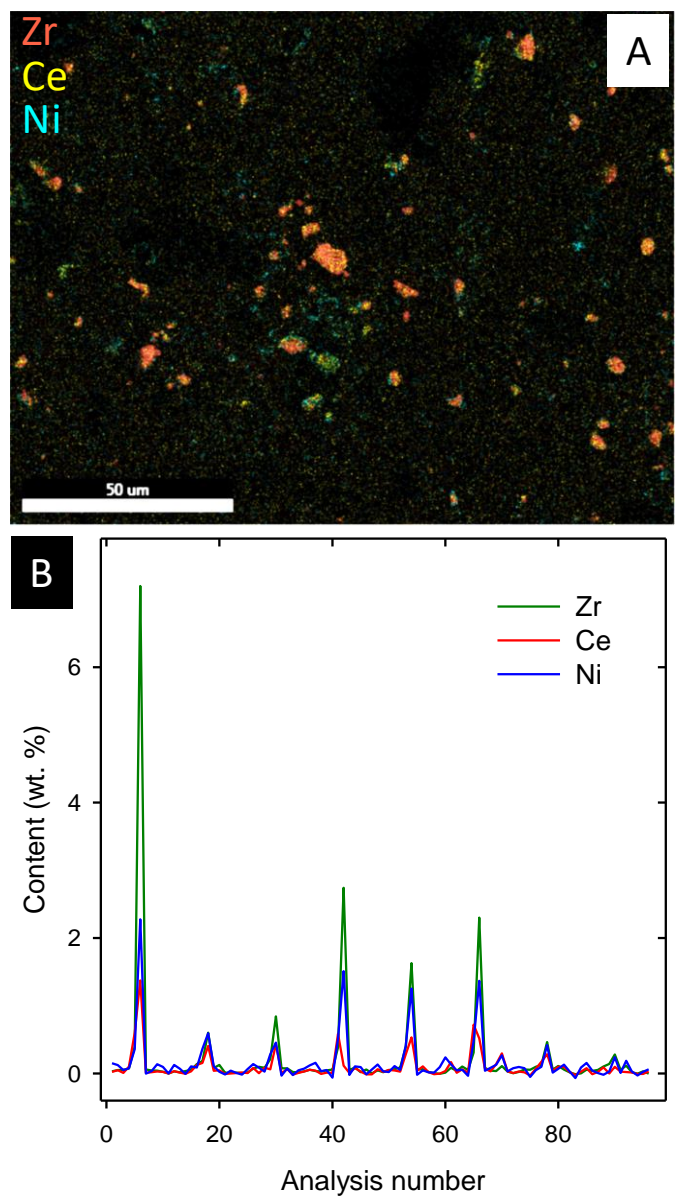

**Fig. 4.** SEM-EDX compositional mapping (A) and micro-XRF analysis (B) representative of the walls surface of the H230NiCZ catalyst.

An electron-transparent cross section of the outer part of the catalyst coating layer, Fig. 5, was cut out by FIB using the standard lift-out technique to investigate in more detail, at the nanometer scale, structural and compositional features of the washcoated layer by TEM/STEM techniques [26]. Fig. 5A shows a HAADF-STEM image from the FIB-lamella: the brightest areas observed at the upper part of the image correspond to the Au and Pt protective layers, intentionally sputtered over the sample during different steps of the cross-section preparation by FIB [51]. A more detailed high-magnification image, Fig. 5B, of the upper part of the lamella provided more information about the coating layer.

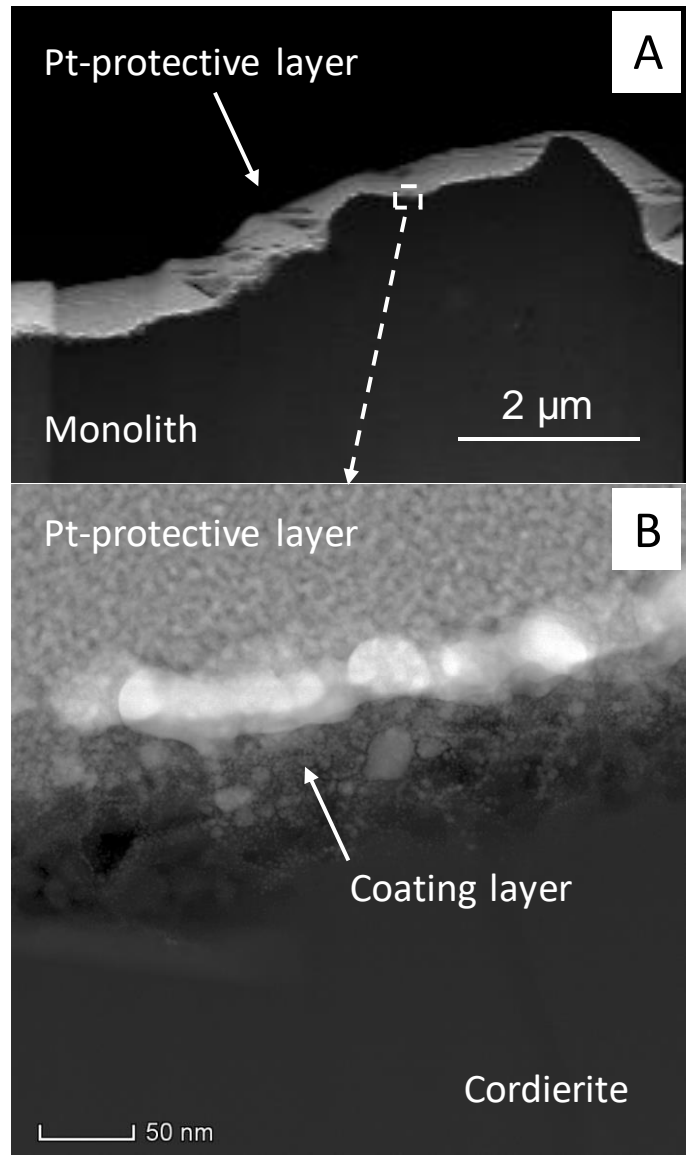

**Fig. 5.** HAADF-STEM images from the H230-NiCZ lamella prepared by FIB at low-magnification (A) and higher magnification image from a selected area (B) revealing the interphase between the coating layer and the cordierite substrate.

In this sense, three different regions can be clearly identified: from the bottom of the image, a highly dense component depicting homogeneous contrast which corresponds to the cordierite substrate, then a second region which seems grainier, that corresponds to the coating layer, and finally, the upper part belonging to the Pt-protective layer with the Au-film sputtered-layer made during the FIB-lamella preparation. From these images, it can be concluded that the coating layer thickness seems to be notably thin, just tens of nanometers ($\approx$ 50 nm). It is remarkable that this nanometric thick washcoat is orders of magnitude thinner than those commonly found in conventional preparations of monoliths [30]. In any case, from the intensity values of these contrasts, it is not possible to discriminate to which component they correspond to. For this purpose, STEM–EDX is the most adequate tool. Note that STEM–EDX maps acquired with nanometer resolution allow discriminating between these components thanks to the elemental chemical analysis. Fig. 6A shows the corresponding chemical results obtained from the same area highlighted in Fig. 5B. Thus, combining EDX chemical elemental maps, the distribution of the different elements comprising the active phase as well as those of the cordierite substrate are now clearly visualized. Comparing the different maps, it seems clear that the highly-dense component, depicting homogeneous contrast, at the bottom part of the HAADF-STEM image, corresponds (see Si elemental map) to the cordierite substrate. Note how the surface of the ceramic monolith is not totally flat but presents local roughness at the scale of tens of nanometers. In fact, a crest and valley structure are clearly revealed both in the HAADF-STEM image and in the Si map.

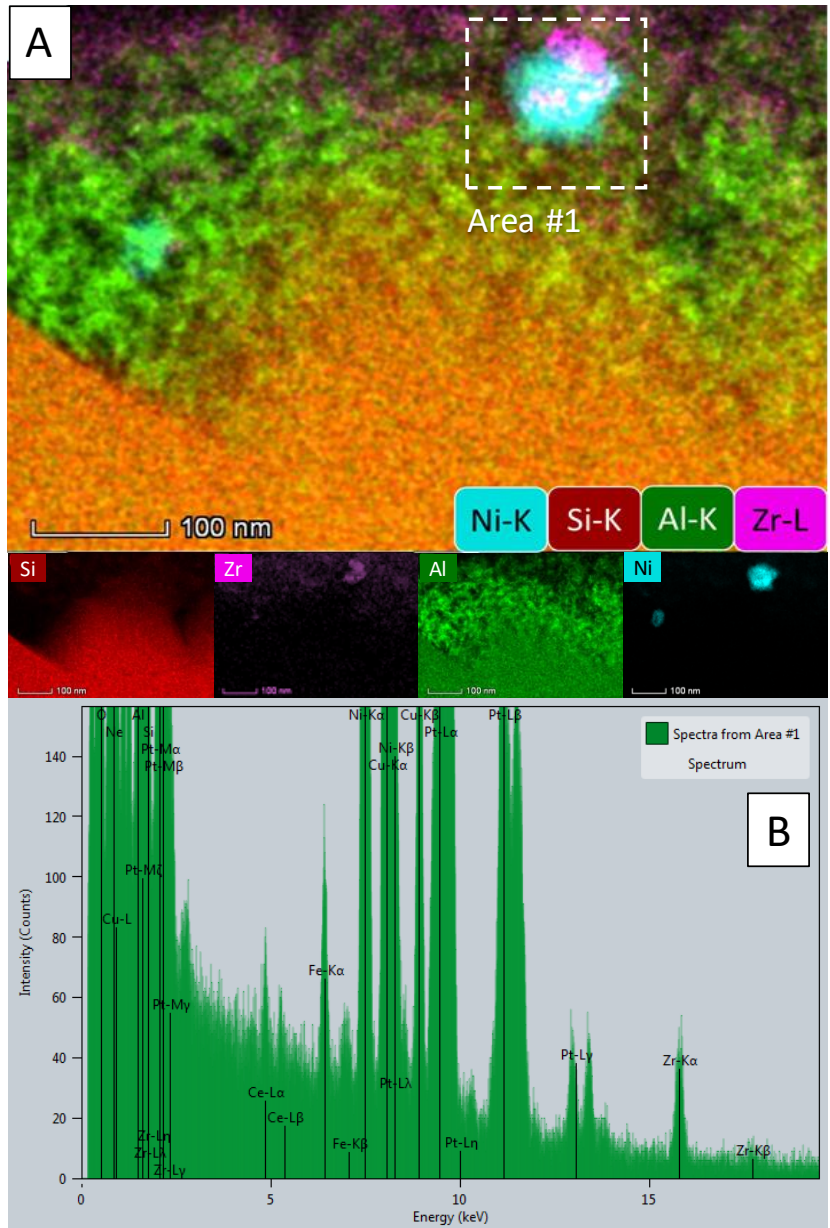

**Fig. 6.** (A) Composed EDX map illustrating the distribution of the different elements from active phase and cordierite substrate. Individual EDX elemental maps for Ni, Al, Si and Zr, respectively are shown below. (B) EDX spectrum from a selected area, labelled as Area#1. All data correspond to a H230-NiCZ lamella prepared by FIB (Fig. 5).

Regarding the Al elemental map, two different regions are observed. One less dense at the top, depicting a grainy aspect, with nanometer-sized pores, where no signal from Si is observed; and a more compact one at the bottom extending over the same area of the Si map. These two different density areas are related with different components.

Thus, the grainier region is assigned to $Al_2O_3$ from the coating layer, since no-Si signal is detected, and the compact one to the cordierite substrate. Note how the coating layer fills perfectly the valleys on the cordierite surface in a continuous fashion, which points outs a successful wetting of the monolith during the washcoating step. Moreover, the coating layer is directly demonstrated to be highly porous, in cross-section, which is undoubtedly a key issue in terms of accessibility of reactants to the active catalyst component.

Lastly, the location of the active phase elements is evidenced by the presence of Ni and Zr signals, respectively. The maps clearly show that Ni nanoparticles are in close contact with Zr-particles. The map corresponding to Ce is not shown in Fig. 6 considering that this lanthanide generates very weak-signals in the EDX experiments, see Fig. 6B, and its elemental maps are much noisier than those of Zr and/or Ni elements. In addition, any elemental quantification should be taken with caution as the analyzed area corresponds to a tiny portion of the washcoat.

Summarizing, the combination of the HAADF-STEM images and the STEM–EDX results have allowed identifying the catalytic active phase nanoparticles in micron-size areas of the coating of our monolithic devices. This suggests that not only the coating layer is extremely thin and porous but also that the active phase is highly spread over the monolith, which is in good agreement with SEM and XRF data as well as with the low loading measured for the devices.

3.3 *Catalytic performance*

In order to establish the operating conditions in the DRM reaction for our catalysts, semi-quantitative Temperature-Programmed Reaction profiles of the evolution of $CO_2$, $CH_4$, $CO$, $H_2$ and $H_2O$ by means of mass spectrometry were recorded (Fig. 7).

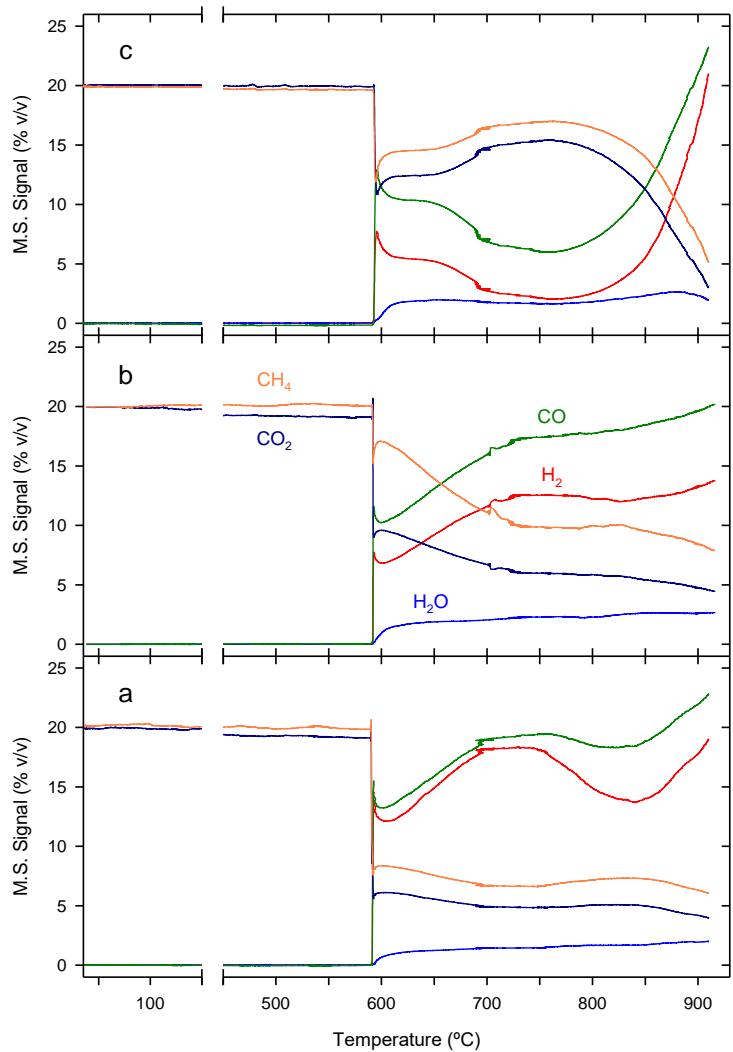

**Fig. 7.** Evolution with temperature of the indicated species for NiCZ (a), NiCZ-S (b) and H230NiCZ (c) samples submitted to heating (5 °C min$^{-1}$) under CH$_4$(20%)/CO$_2$(20%)/He, as recorded by mass spectrometry.

For all the studied samples the reaction abruptly activates around 600 °C, temperature at which CH$_4$ and CO$_2$ consumption along with CO and H$_2$ production are detected. From that point on, the evolution with temperature of the outlet gas composition is different for the NiCZ, NiCZ-S and H230NiCZ samples. As far as the DRM process occurs in parallel and simultaneously with several secondary reactions [52], the detailed analysis of these results becomes complex, in particular because the continuous increase in temperature leads to a non-stationary regime for this type of experiments. Nevertheless, in all the samples, $H_2O$ formation was detected throughout the whole temperature range where catalytic activity is observed, what indicates that the Reverse Water Gas Shift (RWGS) reaction occurs in parallel with DRM. In fact, $CO_2$ conversion values were always higher than those of $CH_4$, and the $H_2/CO$ ratios in the outlet gas were lower than 1. According to the recorded profiles, the temperature range from 700 ºC to 900 ºC would be the most appropriate in relation to the conversion values reached. In this sense, it is important to point out that Wang et al. [53] reported a maximum in the carbon production, caused by $CH_4$ decomposition or the Boudouard reaction, in the temperature range 557 to 700 °C. On the other hand, and also from thermodynamic considerations, temperatures around 900 ºC have been suggested as optima when a $CO_2/CH_4 = 1:1$ feed ratio is employed, since it allows a balance between the conversion achieved and the carbon formation phenomena [5].

Table 2 summarizes the most significant results obtained in the quantitative evaluation of the catalytic activity of the prepared samples in the DRM reaction, monitored by means of gas chromatography. In order to meet stringent conditions in terms of stability, conversion data for $CO_2$ and $CH_4$ along with $H_2/CO$ ratio values after times on stream (TOS) equal or superior to 24 h were measured. As it can be seen, under the studied experimental conditions, even at the highest temperature (900 ºC) the cordierite used as monolithic support is inactive, while the CZ-S sample shows reactant conversions not higher than 10%, similar to those obtained for the H230CZ monolith sample. On the contrary, both honeycomb monolithic catalysts, H230NiCZ and H400NiCZ, exhibit a significant activity in the 700-900 ºC range, the reactant conversion increasing with temperature. As we emphasized in our preliminary study [28], it should be highlighted that this remarkable performance at all temperatures and WHSV values is obtained with monoliths containing a very small specific loading of the washcoat (<0.40 mg cm$^{-2}$), a value at least one order of magnitude lower than in conventional monoliths [49].

Table 2. Catalytic performance in the DRM reaction using a 1:1 $CO_2$:$CH_4$ mixture and after 24 h.

| Sample | WHSV (L g$^{-1}$ h$^{-1}$) | T (°C) | $CO_2$ Conv. (%) | $CH_4$ Conv. (%) | $H_2$/CO ratio |
|---|---|---|---|---|---|
| CZ-S | 115 | 900 | 10 | 8 | 0.15 |
| NiCZ | 115 | 750 | 20 | 11 | 0.25 |
| NiCZ | 115 | 900 | 95 | 92 | 0.89 |
| NiCZ-S | 115 | 750 | 36 | 22 | 0.50 |
| NiCZ-S | 115 | 900 | 84 | 78 | 0.82 |
| H230 [a] | n.a. | 900 | 3 | 0 | - |
| H230CZ | 115 | 900 | 13 | 14 | 0.93 |
| H230NiCZ | 115 | 700 | 84 | 77 | 0.79 |
| H400NiCZ | 115 | 700 | 81 | 73 | 0.78 |
| H230NiCZ [b] | 115 | 750 | 93 | 89 | 0.85 |
| H400NiCZ [c] | 115 | 750 | 89 | 83 | 0.83 |
| H230NiCZ [c] | 231 | 750 | 94 | 90 | 0.86 |
| H230NiCZ | 346 | 750 | 90 | 85 | 0.85 |
| H230NiCZ | 115 | 800 | 93 | 89 | 0.88 |
| H400NiCZ | 115 | 800 | 96 | 93 | 0.90 |
| H230NiCZ | 115 | 900 | 96 | 94 | 0.89 |
| H400NiCZ | 115 | 900 | 99 | 98 | 0.94 |

[a] Data obtained for a complete piece of cordierite; [b] Data after 43 h of reaction; [c] Data after 48 h of reaction.

For these nickel-containing monoliths samples, in the whole temperature interval under study, the $CO_2$ conversion keeps higher while the $CH_4$ conversion is lower than thermodynamic limit values calculated from a direct minimization of Gibbs free energy method using Aspen Plus software [54]. Data reported from these authors (Table 3) consider an equilibrium analysis of a complex multi-reaction system for carbon dioxide reforming of methane in parallel with a significant carbon formation. On the contrary, our results suggest a relatively higher contribution of the RWGS reaction and a less intense methane cracking process when DRM is conducted onto the here proposed monolithic catalysts. Indeed, the calculated thermodynamic conversion for both reactants is closer to the experimental data when the reaction system is modelled minimizing the carbon formation processes (Table 3).

**Table 3.** $CO_2$ and $CH_4$ equilibrium conversion data for the DRM reaction conducted at 1 atm and with $CH_4:CO_2 = 1:1$ based on thermodynamic analysis

| Temperature (ºC) | $CH_4$ Eq. Conv. (%)[a] | $CH_4$ Eq. Conv. (%)[b] | $CO_2$ Eq. Conv. (%)[a] | $CO_2$ Eq. Conv. (%)[b] |
|---|---|---|---|---|
| 700 | 92 | 84 | 65 | 91 |
| 750 | 94 | 91 | 78 | 94 |
| 800 | 96 | 95 | 88 | 97 |
| 900 | 98 | 98 | 96 | 99 |

[a] Analysis reported in [33] based on direct minimization of Gibbs free energy method using Aspen plus sotfware. [b] Data obtained in this work for a reaction system in which the carbon formation processes are minimized using DETCHEM software.

It is noteworthy that both the H230NiCZ and H400NiCZ catalysts present significantly higher conversions than the reference NiCZ and NiCZ-S powders at 750 ºC and 115 L $g^{-1}$ $h^{-1}$, this suggesting that the honeycomb monolithic design (no matter the cell density) prevents from the kinetic control that operates in powdered samples under the highly demanding WHSV regime here employed [55]. This comparison between monolith and powders is also favorable to the first if the activity is expressed as $CH_4$ consumed per unit of time and Ni mass. For example, under the above experimental conditions, we have obtained values of 13.0, 3.2 and 1.6 mmol $CH_4$ $s^{-1}$ $g^{-1}$ Ni, for the H230NiCZ, NiCZ-S and NiCZ samples, respectively, for the TOS values indicated in Table 2. Moreover, it is necessary to increase the temperature of the powdered catalysts up to 900 ºC in order to obtain a catalytic performance similar to that of the monolithic catalysts. Also remarkable, the advantage of the monolithic catalysts remains even at very high flow/catalyst amount ratios, WHSV=346 L $g^{-1}$ $h^{-1}$ (Table 2), unusual experimental conditions scarcely explored in literature [5,25]. In the same way, it is noticeable that our catalysts are very competitive when compared with powdered nickel catalysts onto optimized supports, such as Ni/CeYZrO$_x$ [9], as well as other powdered Ni/MgO-ZrO$_2$ catalysts which contain promoters (Co, Ca, K, Ba, La, Ce) [56], or even with noble metals such as supported Pt or Ru [25]. Moreover, our results are also better when compared with the scarce references that employed Ni catalysts supported onto honeycomb monoliths, either of cordierite [16] or metallic [19]. Furthermore, as mentioned in the introduction section, in the cited references nickel was promoted by small amounts of Ru and Pd-Sn, respectively. Only in [15] some of the samples exhibited superior performance, but the authors employed higher amounts of nickel (11 wt%), which was also doped with alkali and rare-earth metal oxides.

The $H_2$/CO ratio is also an important output in DRM reaction, values close to 1 being desired to allow an easier adaptation to many downstream processes including ammonia and methanol synthesis [25] and Fischer-Tropsch synthesis [6]. On this regard, we obtained $H_2$/CO ratio data ranging from 0.78 up to 0.94 for all the studied nickel-containing honeycomb samples.

Finally, regarding stability under reaction conditions, Fig. 8 shows some reaction profiles versus time selected from the whole series of tested catalysts.

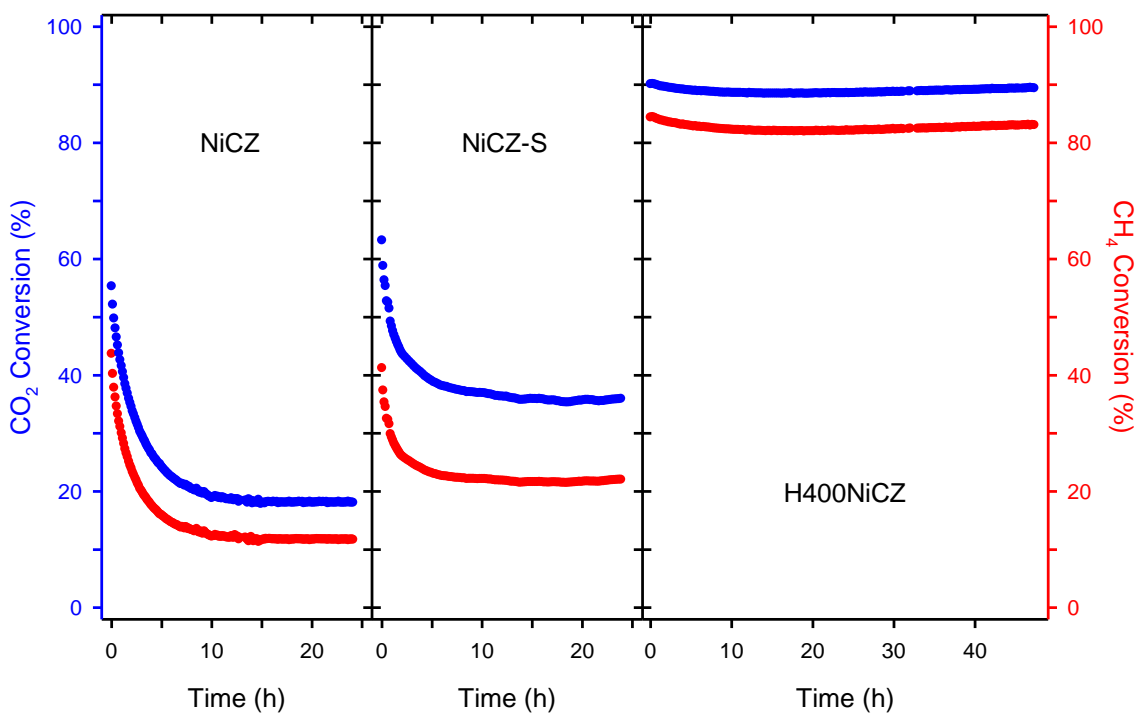

**Fig. 8.** Evolution with TOS of the reactants conversion in the DRM reaction for the catalysts indicated operating at 750 °C with $CH_4:CO_2 = 1:1$ and WHSV = 115 L $g^{-1}$ $h^{-1}$.

Note first that in the case of both powdered catalysts, a clear deactivation phenomenon at 750 ºC is observed, with approx. 70% and 40 % for NiCZ and NiCZ-S, respectively, of conversion decrease after 24 h of TOS. Moreover, this effect is relatively fast being more intense during the first 5 hours. The slightly lower decay in the NiCZ-S sample should be related to the positive effect of alumina on the texture, nickel dispersion and the interaction between the catalyst components above discussed in the characterization results. This interpretation is also supported by literature which demonstrates that a smaller nickel crystal size could retard coking [6,57] while the active phase-alumina interaction might favor resistance against sintering [58,59]. It is particularly remarkable the H400NiCZ catalyst behavior which keeps its high conversion value even after 48 h of reaction time. This result suggests that the honeycomb monolithic design also allows limiting deposition and progressive accumulation of carbon from deep methane cracking reaction [60], which leads to pressure increase in packed beds [54] that in turn may cause flow blockage and even reactor breakup [9]. At this respect, the use of the honeycomb reactor designs could represent a reasonable alternative to conventional catalytic devices or non-conventional technologies recently proposed [61].

## 4. Conclusions

In this work, honeycomb monolithic catalysts of nickel supported on a nanostructured $CeO_2/ZrO_2$ with enhanced redox performance were prepared by the washcoating method, using cordierite with different cell density (230 and 400 cpsi) but reaching similar catalyst specific loading, and tested in the dry reforming of methane (DRM). For the first time, to the best of our knowledge, nickel supported on ceria-zirconia oxides with advanced redox properties, reached by means of severe reduction plus mild oxidation treatments, were employed after deposition onto a honeycomb-type structured support. Working with pure reactants at a relatively low temperature (750 ºC) and operating with WHSV as high as 346 L $g^{-1}$ $h^{-1}$, conversion and $H_2$/CO ratio values close to the thermodynamic limits were reached. Moreover, the catalytic activity kept stable during prolonged experiments of 24-48 h of time-on-stream. These results are particularly noticeable considering that the reference powdered catalyst, studied in parallel, showed poorer catalytic performance, especially for low values of the temperature range studied (750-900 ºC).

STEM analysis of an electron-transparent cross-section of the monolith, prepared by FIB, revealed an ultrathin, just tens of nanometers thick, washcoated layer anchored to the rough surface of the cordierite substrate. This layer is made up of a porous alumina network in which the active phase catalyst nanoparticles are highly spread, thus resulting in very low washcoat loading (<0.40 mg cm$^{-2}$). STEM-EDS analysis also evidenced a thigh contact between nickel, ceria and zirconia in the washcoated layer.

The relative low content and dispersion of nickel in the washcoat does not prevent the catalyst from a very high efficiency in the investigated reaction. This finding can be related with the characterization performed which suggested the occurrence of metal-support interactions. This effect, along with the inherent advantages of the honeycomb monolithic design, might explain the high activity and outstanding stability observed in DRM.

**Author statement**

Fazia Agueniou: Investigation; José M. Gatica: Conceptualization & Methodology, Writing - Review & Editing, Supervision; M. Pilar Yeste: Investigation; Juan C. Hernández-Garrido: Investigation; Miguel A. Cauqui: Conceptualization & Methodology, Funding acquisition; José M. Rodríguez-Izquierdo: Resources; José J. Calvino: Conceptualization & Methodology, Funding acquisition; Hilario Vidal: Conceptualization & Methodology, Writing - Original draft preparation, Supervision.

**Declaration of competing interest**

The authors declare no competing financial interest.

**Acknowledgements**

The authors thank the financial support by the Ministry of Economy and Competitiveness of Spain/FEDER Program of the EU (Project MAT2017-87579-R), and the Junta de Andalucía (Groups FQM-110 and FQM-334). We also acknowledge the Cadiz University SC-ICyT for using its facilities for the ICP-AES, XRD, XRF and SEM-EDX measurements, and the Seville University CITIUS for the FIB-(S)TEM studies.